\documentclass[conference]{IEEEtran}

\IEEEoverridecommandlockouts
\usepackage{cite}
\usepackage{amsmath,amssymb,amsfonts}
\usepackage{algorithmic}
\usepackage{graphicx}
\usepackage{textcomp}
\usepackage{xcolor}
\usepackage{ifpdf}
\usepackage{stfloats}
\usepackage{float}    % 使用float 宏包
\usepackage{hyperref}
\usepackage{subfig}
\usepackage[T1]{fontenc}
\usepackage {indentfirst}
\usepackage{algorithm} 
\usepackage{algorithmic}
\floatname{algorithm}{Algorithm}

\def\BibTeX{{\rm B\kern-.05em{\sc i\kern-.025em b}\kern-.08em
    T\kern-.1667em\lower.7ex\hbox{E}\kern-.125emX}}
\begin{document}

\title{Reconfigurable MIMO towards Electro-magnetic Information Theory: Capacity Maximization Pattern Design}%\\

\author{\IEEEauthorblockN{Haonan Wang\IEEEauthorrefmark{1}, Ang Li\IEEEauthorrefmark{1}, Ya-feng Liu\IEEEauthorrefmark{2}, Qibo Qin\IEEEauthorrefmark{3}, Lingyang Song\IEEEauthorrefmark{4}, and Yonghui Li\IEEEauthorrefmark{5}}
\IEEEauthorblockA{\IEEEauthorrefmark{1}School of Information and Communications Engineering, Xi'an Jiaotong University, Xi'an, Shaanxi, China}
\IEEEauthorblockA{\IEEEauthorrefmark{2}Academy of Mathematics and Systems Science, Chinese Academy of Sciences, Beijing, China}
\IEEEauthorblockA{\IEEEauthorrefmark{3}Wireless Network RAN Research Department, Shanghai Huawei Technologies Co. Ltd., Shanghai, China}
\IEEEauthorblockA{\IEEEauthorrefmark{4}School of Electrical Engineering and Computer Science, Peking University, Beijing, China}% <-this % stops an unwanted space
\IEEEauthorblockA{\IEEEauthorrefmark{5}School of Electrical and Information Engineering, The University of Sydney, Sydney, NSW 2006, Australia}

\IEEEauthorblockN{Email: whn8215858@stu.xjtu.edu.cn\IEEEauthorrefmark{1}, ang.li.2020@xjtu.edu.cn\IEEEauthorrefmark{1}, yafliu@lsec.cc.ac.cn\IEEEauthorrefmark{2}, qinqibo1@huawei.com\IEEEauthorrefmark{3},\\lingyang.song@pku.edu.cn\IEEEauthorrefmark{4}, yonghui.li@sydney.edu.au\IEEEauthorrefmark{5}}
}
\maketitle

\begin{abstract}
In this paper, we focus on the pattern reconfigurable multiple-input multiple-output (PR-MIMO), a technique that has the potential to bridge the gap between electro-magnetics and communications towards the emerging Electro-magnetic Information Theory (EIT). Specifically, we focus on the pattern design problem aimed at maximizing the channel capacity for reconfigurable MIMO communication systems, where we firstly introduce the matrix representation of PR-MIMO and further formulate a pattern design problem. We decompose the pattern design into two steps, i.e., the correlation modification process to optimize the correlation structure of the channel, followed by the power allocation process to improve the channel quality based on the optimized channel structure. For the correlation modification process, we propose a sequential optimization framework with eigenvalue decomposition to obtain near-optimal solutions. For the power allocation process, we provide a closed-form power allocation scheme to redistribute the transmission power among the modified subchannels. Numerical results show that the proposed pattern design scheme offers significant improvements over legacy MIMO systems, which motivates the application of PR-MIMO in wireless communication systems.
\end{abstract}


\section{Introduction}
During the past decades, Multiple-Input Multiple-Output (MIMO) technology has shown its great potential in improving the performance of wireless communication systems because of the ability to exploit the spatial resource over single-antenna systems \cite{1}. 
%More recently, \cite{2} shows that employing reconfigurable antennas at wireless transceivers can provide additional performance gains over traditional antennas with fixed radiation characteristics. Therefore, multi-functional and reconfigurable MIMO (PR-MIMO) techniques have gained increasing research attention in recent years \cite{3}.
Nevertheless, the capacity of modern MIMO communication systems have shown to approach the Shannon limits. To meet the ever-increasing demand for data rates in 5.5G/6G and beyond, the concept of Electromagnetic Information Theory (EIT) has recently been proposed \cite{38}, which aims to merge the electro-magnetics and information theory that have been studied separately for years. Pattern reconfigurable MIMO (PR-MIMO) based on reconfigurable antennas is able to affect the electro-magnetic fields via reconfiguring the radiation pattern, which has the potential to bridge the gap between electro-magnetics and information theory \cite{3}.

%Reconfigurable antennas form a special class of antennas which can configure different radiation characteristics such as frequency bands, polarizations or radiation patterns via electrical and
Based on the advances on antenna design techniques, reconfigurable antennas can be designed to operate with different radiation characteristics, such as frequency bands, polarizations or radiation patterns \cite{4,5}.
%Reconfigurable antennas form a special class of antennas which can be configured to operate with different frequency bands, different polarizations or radiation patterns \cite{4,5}.
%Among different types of reconfigurable antennas, the pattern reconfigurability, which is the focus of this paper, can further enhance the ability of interference cancellation and resource allocation by improving the degrees of freedom (DoFs) in the signal directions.
Compared with other reconfigurable dimensions, the pattern reconfigurability, which is mainly discussed in this paper, can further enhance the ability of interference cancellation and resource allocation by improving the degrees of freedom (DoFs).
%can improve the degrees of freedom (DoFs) in the signal directions so that the ability of interference suppression and data transmission can be further enhanced.
PR-MIMO can offer spatial diversity in radiation directivity at the same transmission frequency, and 
%a considerable number of methods to realize PR-MIMO have already been promoted \cite{10},
%there already existed a considerable number of methods to realize PR-MIMO \cite{10}, 
a great quantity of methods have been studied to realize PR-MIMO \cite{10},
such as switching the circuit \cite{11,12}, choosing different radiation units, etc. \cite{16,13,14}.
%A common method was to use a switching circuit to feed a sector array such that only one array element was executed at each timeslot \cite{11,12}. Meanwhile, other methods for realizing pattern reconfiguration include choosing different radiation units, changing the characteristic modes of radiators, etc., \cite{16,17}. %Because of the ability to increase the signal transmission distance and quality, pattern reconfigurable MIMO has been mostly promising in the region of surveillance and tracking. For example, \cite{18} designed a high-gain pattern reconfigurable MIMO antenna array to increase the power efficiency in wireless handheld terminals. What's more, \cite{19} proposed an approach to analyze the characterization of pattern reconfigurable antennas designed for MIMO systems. It revealed that pattern reconfigurable MIMO can redirect the signal to intended users so that the energy efficiency can be increased and the communication coverage can be extended. 

%The performance benefits of PR-MIMO mainly come from the additional DoFs, and the exploitation of such additional DoFs is mainly accomplished through the effective optimal mode selection scheme. 

%Currently, several applications of PR-MIMO have been promoted in areas such as MIMO transmission \cite{20}, target detection and tracking \cite{21}, direction of arrival (DoA) estimation \cite{22,23}. 
The applications of PR-MIMO in practical wireless scenarios such as user scheduling \cite{20}, target location \cite{21}, direction of arrival (DoA) estimation \cite{22,23}, have been discussed recently.
%In \cite{20}, the additional efficient channels generated by PR-MIMO improved the performance by expanding the users scheduling region. 
In \cite{20}, PR-MIMO improved the performance by expanding the users scheduling region via the additional reconfigurable channels.
The pairing scheduling was determined by the proposed joint user and antenna mode selection scheme, and a greedy low-complexity iterative selection algorithm was further designed. \cite{21} considered PR-MIMO for target detection and tracking, where a Bayesian cognitive target tracking technique to minimize the Cramér-Rao lower bound (CRLB) of the DoA parameters was proposed. Considering the DoA estimation for PR-MIMO, \cite{22} and \cite{23} combined the traditional methods with the adaptive decision algorithm and further improved the estimation accuracy.
%With respect to the DoA estimation for PR-MIMO, the traditional estimation technique was combined with the mode selection scheme so that an improved performance could be achieved \cite{22,23}.

\begin{figure*}[ht]
	\centering
	\includegraphics[width=6.5in]{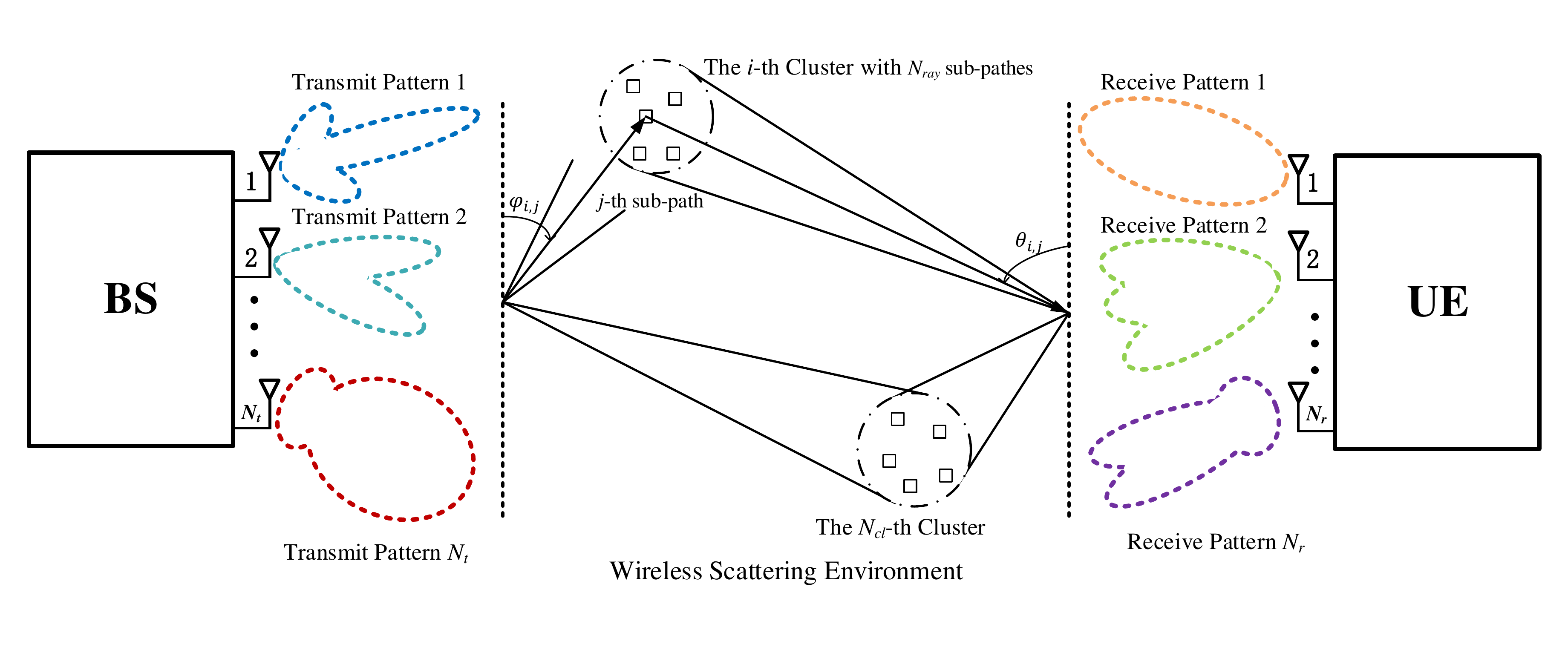}
	\caption{Multi-path channel model for PR-MIMO.}
	\label{structure}
\end{figure*}

However, two challenges that hinder the widely application of PR-MIMO still exist. On one hand, the exploitation of the additional DoFs has to be accomplished through the effective optimal mode selection, which brings considerable channel estimation overhead. The channel extrapolation based on the pattern correlation \cite{24} and the optimal decision algorithm based on Multi-Armed Bandit (MAB) \cite{25,26,27,28,29} are the major 
attempts to solve this problem currently.
Moreover, the physical mechanism of the modification process and the quality improvement in PR-MIMO has not been revealed, and the methodology of the radiation pattern design in PR-MIMO systems for optimizing some wireless performance metrics such as capacity maximization has not been established.
%it is not clear how to design the optimal radiation pattern for capacity maximization in PR-MIMO systems, which is the focus of this paper.

In this paper, we study the pattern design problem aimed at maximizing the channel capacity for PR-MIMO systems. We firstly %introduce the matrix representation framework of PR-MIMO 
describe the pattern channel in a matrix form
and further formulate a pattern design optimization problem. After revealing the physical mechanism of the pattern effect on the scattering channel, we decompose the optimal pattern design into two steps, i.e., the correlation modification process to optimize the correlation structure of the channel, and the power allocation process to improve the channel quality, sequentially. %For the correlation modification process, we propose a sequential optimization framework (SOF) to transform the design of the correlation modification matrix into a sequential optimization problems on correlation modification vectors, where a criterion to determine their sequences to be optimized is presented. 
For the correlation modification process, we propose a sequential optimization framework (SOF) to transform the correlation modification matrix design problem into a sequential vector optimization problems, which can be solved using eigenvalue decomposition based method, and give a criterion to determine their optimized sequences.
%The subproblem for each vector optimization is solved via eigenvalue decomposition.
For the power allocation process, a closed-form scheme is designed to redistribute the transmission power according to the independence level of the subchannels. Numerical results demonstrate the efficient channel modification ability of the designed reconfigurable pattern and further validate the significant improvements of PR-MIMO over legacy MIMO systems.

Notations: $a$, $\boldsymbol{a}$, and $\mathbf{A}$ denote scalar, column vector and matrix, respectively. $(\cdot)^{*}$, $(\cdot)^{\mathrm{T}}$, and $(\cdot)^{\mathrm{H}}$ denote conjugate, transposition, conjugate transposition respectively while $\operatorname{Tr}\left(\cdot\right)$ represents the trace of a matrix. $\operatorname{vec}\left(\cdot\right)$ denotes the vectorization operator and $\operatorname{diag}\left(\cdot\right)$ denotes the diagonalization operator. $\left\langle\cdot\right\rangle$ denotes the inner product operation, i.e., $\left\langle\boldsymbol{a},\boldsymbol{b}\right\rangle=\boldsymbol{a}^{\mathrm{H}}\boldsymbol{b}$ for complex vectors $\boldsymbol{a}$ and $\boldsymbol{b}$. Frobenius norm of a matrix is denoted by $\|\cdot\|_{\mathrm{F}}$. $\mathbf{I}_{M}$ and $\mathbf{1}_{M \times N}$ represent a $M \times M$ identity matrix and a $M \times N$ with all entries being $1$. $\mathcal{S}_L$ denotes all $L \times L$ symmetric matrices. $\boldsymbol{e}_i$ is the $i$-th column of an identity matrix. $\succeq$ and $\preceq$ represent matrix generalized inequalities. $\odot$ denotes Hadamard product.
\section{System Model}

%In this section, we firstly present the preliminaries of the pattern reconfigurable antenna array. Subsequently, the multi-path channel model and the channel model incorporating pattern reconfigurable antenna array are introduced, respectively.
In this section, we firstly present the multi-path channel model and %the channel model incorporating pattern reconfigurable antenna array. 
the pattern reconfigurable channel model.
After that, the capacity of PR-MIMO is introduced.

%For the pattern reconfigurable antenna array, we introduce $\vec{f}(\theta, \phi, \mu)$ to denote the complex far-field radiation pattern, where $\theta$ and $\phi$ denote the corresponding vertical and horizontal angle, and $\mu$ is the index of the antenna pattern mode. The spherical coordinate representation of the complex electric field for the $m$-th antenna element can be expressed as 
%\begin{equation}
%	\vec{f}\left(\theta, \phi, \mu_{m}\right)=f_{\theta}\left(\theta, \phi, \mu_{m}\right) \vec{e}_{\theta}+f_{\phi}\left(\theta, \phi, \mu_{m}\right) \vec{e}_{\phi},
%\end{equation}
%where $\vec{e}_{\theta}$ and $\vec{e}_{\phi}$ denote the unit vectors in the $\theta$ and $\phi$ direction, respectively.
\subsection{Multi-Path MIMO Channel Model}

We consider a point-to-point (P2P) MIMO system in the downlink, 
%where the number of transmit and receive antennas is $N_t$ and $N_r$ 
equipped with $N_t$ antennas at the transmitter and $N_r$ antennas at the receiver, where $N_r \leq N_t$, as shown in Fig. 1. Assuming the uniform linear array (ULA) at the transceivers, the physical multi-path MIMO channel can be modeled as \cite{7}:
\begin{equation}
	\mathbf{H}=\sum_{l=1}^{L}\alpha_{l}\boldsymbol{a}_{\mathrm{R}}\left(\theta_{l}\right)\boldsymbol{a}_{\mathrm{T}}^{\mathrm{H}}\left(\varphi_{l}\right),
\end{equation}
where $\mathbf{H} \in \mathbb{C}^{N_r \times N_t}$ represents the channel matrix with the power constraint $\mathbb{E}\left[\|\mathbf{H}\|_{\mathrm{F}}^{2}\right]=N_{t} N_{r}$, $L$ is the scattering paths number and
%with $L=N_{\mathrm{cl}}\times N_{\mathrm{ray}}$ where $N_{\mathrm{cl}}$ denotes the number of scattering clusters and $N_{\mathrm{ray}}$ denotes the number of scattering rays in each cluster, 
$\alpha_l$ denotes the channel gain of the $l$-th scattering path. %In (1), $\boldsymbol{a}_{\mathrm{R}}\left(\theta_{l}\right)$ and $\boldsymbol{a}_{\mathrm{T}}\left(\varphi_{l}\right)$ denote the receive and transmit array response vectors, where $\theta_{l}$ and $\varphi_{l}$ represent the azimuth angles of arrival and departure (AoAs and AoDs), respectively. The array response vectors are given by 
The steering vectors at the receiver and the transmitter are given by
\begin{equation}
	\boldsymbol{a}_{\mathrm{R}}\left(\theta_{l}\right)=\dfrac{1}{\sqrt{N_r}} \left[1, e^{-j 2 \pi \frac{d_{\mathrm{R}}}{\lambda} \sin \theta_{l}}, \ldots, e^{-j 2 \pi \frac{d_{\mathrm{R}}}{\lambda}\left(N_{r}-1\right) \sin \theta_{l}}\right]^{\mathrm{T}}
\end{equation}
and
\begin{equation}
	\boldsymbol{a}_{\mathrm{T}}\left(\varphi_{l}\right)=\dfrac{1}{\sqrt{N_t}} \left[1, e^{-j 2 \pi \frac{d_{\mathrm{T}}}{\lambda} \sin \varphi_{l}}, \ldots, e^{-j 2 \pi \frac{d_{\mathrm{T}}}{\lambda}\left(N_{t}-1\right) \sin \varphi_{l}}\right]^{\mathrm{T}},
\end{equation}
where $\frac{d_{\mathrm{T}}}{\lambda}$ and $\frac{d_{\mathrm{R}}}{\lambda}$ represent the normalized antenna spacing at the transceivers, with the carrier wavelength $\lambda$. $\theta_{l}$ and $\varphi_{l}$ denote the azimuth angles of arrival and departure (AoAs and AoDs). Based on that, (1) can be simplified into
\begin{equation}
	\mathbf{H}=\mathbf{A}_{\mathrm{R}} \mathbf{\Lambda} \mathbf{A}_{\mathrm{T}}^{\mathrm{H}},
\end{equation}
where
\begin{equation}
	\mathbf{A}_{\mathrm{R}}=\left[\boldsymbol{a}_{\mathrm{R}}\left(\theta_{1}\right), \boldsymbol{a}_{\mathrm{R}}\left(\theta_{2}\right), \ldots, \boldsymbol{a}_{\mathrm{R}}\left(\theta_{L}\right)\right]
\end{equation}
and
\begin{equation}
	\mathbf{A}_{\mathrm{T}}=\left[\boldsymbol{a}_{\mathrm{T}}\left(\varphi_{1}\right), \boldsymbol{a}_{\mathrm{T}}\left(\varphi_{2}\right), \ldots, \boldsymbol{a}_{\mathrm{T}}\left(\varphi_{L}\right)\right]
\end{equation}
are the steering matrices of the transceivers, and $\mathbf{\Lambda}=\operatorname{diag}\left\{\alpha_{1}, \alpha_{2}, \ldots, \alpha_{L}\right\}$ contains the complex channel gains on the diagonal.

%Recalling that the above channel model $\mathbf{H}$ only describes the physical wireless propagation environments without considering the antenna radiation pattern, we call it {\it physical channel} in the following part.
Considering that the channel model $\mathbf{H}$ only describes the physical scattering information without the antenna radiation characteristics, we term it the {\it physical channel}.
%describes the physical wireless scattering environments without considering the antenna radiation pattern. For convenience, we call it {\it physical channel} in the following part.

\subsection{PR-MIMO Channel Model}
\newcounter{TempEqCnt}                         % 创建临时变量TempEqCnt
\setcounter{TempEqCnt}{\value{equation}} % 将当前公式序号 赋给TempEqCnt
\setcounter{equation}{8}                           % 当前公式序号变为x，x等于长公式应有的序号减1.
\begin{figure*}[hb]
	\hrulefill
	\begin{equation}
		\begin{aligned}
			\widetilde{\mathbf{H}}= & \sum_{l=1}^{L} \alpha_{l} \boldsymbol{a}_{\mathrm{R}}\left(\theta_{l}\right) \boldsymbol{a}_{\mathrm{T}}^{\mathrm{H}}\left(\varphi_{l}\right) \odot\left({\left[\begin{array}{cl}
					F_{\mathrm{v}, \theta_{l}, v_{1}}^{\mathrm{Re}} & F_{\mathrm{h}, \vartheta_{l}, v_{1}}^{\mathrm{Re}}  \\
					F_{\mathrm{v}, \theta_{l}, v_{2}}^{\mathrm{Re}} & F_{\mathrm{h}, \vartheta_{l}, v_{2}}^{\mathrm{Re}}  \\
					\vdots & \vdots  \\
					F_{\mathrm{v}, \theta_{l}, v_{N_{r}}}^{\mathrm{Re}} & F_{\mathrm{h}, \vartheta_{l}, v_{N_{r}}}^{\mathrm{Re}}
				\end{array}\right]}  \left[\begin{array}{llll}
					F_{\mathrm{v}, \varphi_{l}, \mu_{1}}^{\mathrm{Tr}} & F_{\mathrm{v}, \varphi_{l}, \mu_{2}}^{\mathrm{Tr}} & \ldots & F_{\mathrm{v}, \varphi_{l}, \mu_{N_{t}}}^{\mathrm{Tr}} \\
					\\
					F_{\mathrm{h}, \phi_{l}, \mu_{1}}^{\mathrm{Tr}} & F_{\mathrm{h}, \phi_{l}, \mu_{2}}^{\mathrm{Tr}} & \ldots & F_{\mathrm{h}, \phi_{l}, \mu_{N_{t}}}^{\mathrm{Tr}}
				\end{array}\right]\right).
		\end{aligned}
	\end{equation}
\end{figure*}
\setcounter{equation}{\value{TempEqCnt}}

\setcounter{TempEqCnt}{\value{equation}} % 将当前公式序号 赋给TempEqCnt
\setcounter{equation}{10}                           % 当前公式序号变为x，x等于长公式应有的序号减1.
\begin{figure*}[hb]
	\hrulefill
	\begin{equation}
		\begin{aligned}
			\widetilde{\mathbf{H}}= & \sum_{l=1}^{L} \alpha_{l} \boldsymbol{a}_{\mathrm{R}}\left(\theta_{l}\right) \boldsymbol{a}_{\mathrm{T}}^{\mathrm{H}}\left(\varphi_{l}\right) \odot\left[\begin{array}{cccc}
			F_{\mathrm{v}, \varphi_{l}, \mu_{1}}^{\mathrm{Tr}} & F_{\mathrm{v}, \varphi_{l}, \mu_{2}}^{\mathrm{Tr}} &  \cdots & F_{\mathrm{v}, \varphi_{l}, \mu_{N_{t}}}^{\mathrm{Tr}}  \\
			F_{\mathrm{v}, \varphi_{l}, \mu_{1}}^{\mathrm{Tr}} & F_{\mathrm{v}, \varphi_{l}, \mu_{2}}^{\mathrm{Tr}} &  \cdots & F_{\mathrm{v}, \varphi_{l}, \mu_{N_{t}}}^{\mathrm{Tr}}  \\
			\vdots  & \ddots & \vdots & \vdots \\
			F_{\mathrm{v}, \varphi_{l}, \mu_{1}}^{\mathrm{Tr}} & F_{\mathrm{v}, \varphi_{l}, \mu_{2}}^{\mathrm{Tr}} &  \cdots & F_{\mathrm{v}, \varphi_{l}, \mu_{N_{t}}}^{\mathrm{Tr}} 
		\end{array}\right]\\
		= & \mathbf{A}_{\mathrm{R}}\mathbf{\Lambda}\left(\mathbf{A}_{\mathrm{T}}\odot\left[\begin{array}{cccc}
			F_{\mathrm{v}, \varphi_{1}, \mu_{1}}^{\mathrm{Tr}} & F_{\mathrm{v}, \varphi_{2}, \mu_{1}}^{\mathrm{Tr}} &  \cdots & F_{\mathrm{v}, \varphi_{L}, \mu_1}^{\mathrm{Tr}}  \\
			F_{\mathrm{v}, \varphi_{1}, \mu_{2}}^{\mathrm{Tr}} & F_{\mathrm{v}, \varphi_{2}, \mu_{2}}^{\mathrm{Tr}} &  \cdots & F_{\mathrm{v}, \varphi_{L}, \mu_{2}}^{\mathrm{Tr}}  \\
			\vdots  & \ddots & \vdots & \vdots \\
			F_{\mathrm{v}, \varphi_{1}, \mu_{N_{t}}}^{\mathrm{Tr}} & F_{\mathrm{v}, \varphi_{2}, \mu_{N_{t}}}^{\mathrm{Tr}} &  \cdots & F_{\mathrm{v}, \varphi_{L}, \mu_{N_{t}}}^{\mathrm{Tr}} 
		\end{array}\right]\right)^{\mathrm{H}},\\
		= & \mathbf{A}_{\mathrm{R}}\mathbf{\Lambda}\left(\mathbf{A}_{\mathrm{T}}\odot\mathbf{M}\right)^{\mathrm{H}}.
		\end{aligned}
	\end{equation}
\end{figure*}
\setcounter{equation}{\value{TempEqCnt}}

Considering pattern reconfigurable antennas at both the transceivers, 
%the element-wise pattern reconfigurable MIMO channel model is represented as \cite{24}
the complex channel gain between the $m$-th transmit antenna and the $n$-th receive antenna is represented as \cite{24}
\begin{equation}
	\begin{aligned}
		\widetilde{\mathbf{H}}_{n, m}\left(\mu_m, v_n\right)=\sum_{l=1}^{L} \alpha_{l} & \left\langle\vec{f}_{\mathrm{T}}\left(\varphi_{l}, \phi_{l}, \mu_m\right), \vec{f}_{\mathrm{R}}\left(\theta_{l}, \vartheta_{l}, v_n\right)\right\rangle \\
		& e^{j 2 \pi\left(\frac{d_{\mathrm{T}}}{\lambda}(m-1) \sin \varphi_{l}-\frac{d_{\mathrm{R}}}{\lambda}(n-1) \sin \theta_{l}\right)},
	\end{aligned}
\end{equation}
%where $\widetilde{\mathbf{H}}_{n, m}\left(\mu_m, v_n\right)$ denotes the channel between the $m$-th transmit antenna with mode $\mu_m$ and the $n$-th receive antenna with mode $v_n$, while $\mu_m \in \mathcal{R}_m$ and $v_n \in \mathcal{R}_n$ where $\mathcal{R}_m$ and $\mathcal{R}_n$ denote the continuous radiation pattern mode sets of the $m$-th transmit antenna and the $n$-th receive antenna, 
where $\vec{f}_{\mathrm{T}}\left(\varphi_{l}, \phi_{l}, \mu_m\right)=\left[F_{\mathrm{v}, \varphi_{l}, \mu_m}^{\mathrm{Tr}}, F_{\mathrm{h}, \phi_{l}, \mu_m}^{\mathrm{Tr}}\right]^{\mathrm{T}}$ and $\vec{f}_{\mathrm{R}}\left(\theta_{l}, \vartheta_{l}, v_n\right)=\left[F_{\mathrm{v}, \theta_{l}, v_n}^{\mathrm{Re}}, F_{\mathrm{h}, \vartheta_{l}, v_n}^{\mathrm{Re}}\right]^{\mathrm{T}}$ are the electric pattern vectors of
the $m$-th transmit antenna with pattern $\mu_m$ and the $n$-th receive antenna with pattern $v_n$
in the direction of $\mathrm{AoD}=\varphi_{l}$, $\mathrm{EoD}=\phi_{l}$, $\mathrm{AoA}=\theta_{l}$ and $\mathrm{EoA}=\vartheta_{l}$, where $\mathrm{EoD}$ and $\mathrm{EoA}$ are abbreviations for elevation angle of departure and that of arrival, while $\mathrm{AoD}$ and $\mathrm{AoA}$ are abbreviations for azimuth angle of departure and that of arrival, respectively. $F_{\mathrm{v}}$ and $F_{\mathrm{h}}$ denote the projection of the pattern vector on the vertical and horizontal dimensions.
%vertical and horizontal components of the pattern vector in spherical coordinates. %Compared with $(2)$, the effect of the antenna pattern is described by the inner product of the corresponding pattern vectors. 
According to (7), the inner product of the pattern vectors describes the modification process of the antenna pattern.
Similarly, we term $\widetilde{\mathbf{H}}$ the {\it pattern channel} matrix.

%The transmission process of the PR-MIMO system can be written as
%\begin{equation}
%	\mathbf{y}=\widetilde{\mathbf{H}}\mathbf{x}+\mathbf{n},
%\end{equation}
%where $\mathbf{x}$ and $\mathbf{y}$ represent transmit and receive signal vectors, and $\mathbf{n}\sim \mathcal{CN}\left(0, \sigma_{n}^{2}\mathbf{I}_{N_r}\right)$ is the additive white Gaussian noise (AWGN) vector at the receiver side.

\setcounter{equation}{11}                           % 当前公式序号变为x，x等于长公式应有的序号减1.
\begin{figure*}[hb]
	\hrulefill
	\begin{equation}
		\begin{aligned}
			\mathcal{P}_{1}: \max _{\mathbf{M}} \ & \log _{2} \operatorname{det}\left({\mathbf{I}_{N_{r}}+\frac{\rho}{N_{r}} \mathbf{A}_{\mathrm{R}} \mathbf{\Lambda}\left(\mathbf{A}_{\mathrm{T}} \odot \mathbf{M}\right)^{\mathrm{H}}} {\left(\mathbf{A}_{\mathrm{T}} \odot \mathbf{M}\right) \mathbf{\Lambda}^{\mathrm{H}} \mathbf{A}_{\mathrm{R}}^{\mathrm{H}}}\right) \\
			\text { s.t. } & \operatorname{Tr}\left(\mathbf{A}_{\mathrm{R}} \mathbf{\Lambda}\left(\mathbf{A}_{\mathrm{T}} \odot \mathbf{M}\right)^{\mathrm{H}}\left(\mathbf{A}_{\mathrm{T}} \odot \mathbf{M}\right) \mathbf{\Lambda}^{\mathrm{H}} \mathbf{A}_{\mathrm{R}}^{\mathrm{H}}\right) \leq N_{r} N_{t}, \\
			& \mathbf{M}_{i, j} \geq 0 \quad i, j=1,2, \ldots, L.
		\end{aligned}
	\end{equation}
\end{figure*}
\setcounter{equation}{\value{TempEqCnt}}

\setcounter{TempEqCnt}{\value{equation}} % 将当前公式序号 赋给TempEqCnt
\setcounter{equation}{15}                           % 当前公式序号变为x，x等于长公式应有的序号减1.
\begin{figure*}[hb]
	\hrulefill
	\begin{equation}
		\begin{aligned}
			\widehat{\mathbf{G}}_{i, j} &=\operatorname{Tr}\left(\widehat{\mathbf{H}}_{i}^{\mathrm{H}} \widehat{\mathbf{H}}_{j}\right) \\
			&=\operatorname{Tr}\left(\left(\boldsymbol{a}_{\mathrm{T}, i} \odot \widehat{\boldsymbol{m}}_{i}\right) \boldsymbol{a}_{\mathrm{R}, i}^{\mathrm{H}} \boldsymbol{a}_{\mathrm{R}, j}\left(\boldsymbol{a}_{\mathrm{T}, j} \odot \widehat{\boldsymbol{m}}_{j}\right)^{\mathrm{H}}\right) \\
			&=\frac{1}{N_r N_t}\sum_{n=1}^{N_{r}} e^{j 2 \pi \frac{d_{\mathrm{R}}}{\lambda}(n-1)\left(\sin \theta_{j}-\sin \theta_{i}\right)}  \sum_{k=1}^{N_{t}} \widehat{\boldsymbol{m}}_{i}(k) \widehat{\boldsymbol{m}}_{j}(k) e^{j 2 \pi \frac{d_{\mathrm{T}}}{\lambda}(k-1)\left(\sin \varphi_{i}-\sin \varphi_{j}\right)}.
		\end{aligned}
	\end{equation}
\end{figure*}
\setcounter{equation}{\value{TempEqCnt}}

\subsection{Capacity of PR-MIMO}
Assuming that the transmitter holds the perfect channel state information, the capacity for a PR-MIMO system is given by
\setcounter{equation}{7}      
\begin{equation}
	C=\log _{2} \operatorname{det}\left(\mathbf{I}_{N_{r}}+\frac{\rho}{N_{r}} \widetilde{\mathbf{H}} \widetilde{\mathbf{H}}^{\mathrm{H}}\right),
\end{equation}
where $\rho$ represents the transmit signal-to-noise ratio (SNR).

%\section{Capacity Maximized Pattern Design}
\section{Problem Formulation}
%In this section, the matrix representation of the pattern channel is proposed, based on which the pattern design problem is further formulated.
In this section, we firstly discuss how to describe the pattern channel in a matrix form, based on which the pattern design problem is further formulated.

\subsection{Matrix Representation of PR-MIMO}

Firstly, the reason to describe the pattern channel in a matrix form should be explained. 
%At present, the pattern channel is represented in (7), which only describes the generation process of each element in the channel matrix. However, such an expression does not fully reveal the effect of the pattern, and meanwhile brings difficulties to the optimal pattern design.
The element-wise channel representation in (7) only describes the mathematical generation process of the pattern channel, but cannot reveal the physical mechanism of the pattern effect. The optimal pattern design calls for a matrix representation of the PA-MIMO channel.

%The description in (7) can be extended to a matrix form, as shown in (9) at the bottom of this page. 
The expansion of the matrix form is shown in (9) at the bottom of this page. Considering the transmit pattern reconfigurability only in this paper for simplicity, the receiver pattern is given by

\setcounter{equation}{9}
\begin{equation}
    \left[\begin{array}{cc}
		F_{\mathrm{v}, \theta_{l}, v_{1}}^{\mathrm{Re}} & F_{\mathrm{h}, \vartheta_{l}, v_{1}}^{\mathrm{Re}}  \\
		F_{\mathrm{v}, \theta_{l}, v_{2}}^{\mathrm{Re}} & F_{\mathrm{h}, \vartheta_{l}, v_{2}}^{\mathrm{Re}}  \\
		\vdots & \vdots  \\
		F_{\mathrm{v}, \theta_{l}, v_{N_{r}}}^{\mathrm{Re}} & F_{\mathrm{h}, \vartheta_{l}, v_{N_{r}}}^{\mathrm{Re}}
	\end{array}\right]=\left[\begin{array}{cc}
		1 & 0 \\
		1 & 0 \\
		\vdots & \vdots \\
		1 & 0
	\end{array}\right].
\end{equation}
where only the vertical polarization is considered. Based on that, (9) can be further simplified into (11) at the bottom of this page. 

According to (11), the reconfigurable pattern modifies the channel via introducing an extra power factor in the corresponding transmission direction.
%the influence of the pattern on the channel can be regarded as an additional power gain in the corresponding propagation direction. 
The pattern sampling matrix $\mathbf{M}$ in (11), whose $(k,l)$-th element denotes the sampling radiation pattern gain of the $k$-th antenna element in the $l$-th scattering direction, reveals the physical mechanism of pattern modification and motivates us to formulate the pattern design problem, as shown in the following.

\subsection{Problem Formulation}
%In this paper, we mainly discuss the pattern design to maximize the capacity of the PR-MIMO system, which 

The capacity maximization pattern design problem can be constructed as $\mathcal{P}_1$ in (12) at the bottom of this page. The first constraint prevents the additional power gain, while the second one enforces that only the power modification effect of the pattern is considered, without the phase adjustment ability. Considering that all the antenna elements are equipped with the same reconfigurable pattern such that $\operatorname{rank}\left(\mathbf{M}\right)=1$, 
%Considering the condition when $\operatorname{rank}\left(\mathbf{M}\right)=1$,
$\mathcal{P}_1$ can be further transformed into
\setcounter{equation}{12}
\begin{equation}
	\begin{aligned}
		\mathcal{P}_{2}: \min _{\mathbf{X}} \ &-\log _{2} \operatorname{det}\left(\mathbf{I}_{N_{r}}+\frac{\rho}{N_{r}} \mathbf{A}_{\mathrm{R}} \mathbf{\Lambda}\left(\mathbf{R}_{\mathrm{T}} \odot \mathbf{X}\right) \mathbf{\Lambda}^{\mathrm{H}} \mathbf{A}_{\mathrm{R}}^{\mathrm{H}}\right) \\
		\text { s.t. } &\operatorname{Tr}\left(\mathbf{A}_{\mathrm{R}} \mathbf{\Lambda}\left(\mathbf{R}_{\mathrm{T}} \odot \mathbf{X}\right) \mathbf{\Lambda}^{\mathrm{H}} \mathbf{A}_{\mathrm{R}}^{\mathrm{H}}\right) \leq  N_{r} N_{t}, \\
	%	\quad&-\mathbf{X}_{i, j} \leq 0 \\
	%	&\mathbf{X}=\mathbf{X}^{\mathrm{T}} \\
	%	& \mathbf{X}\succeq 0 \\
		& \mathbf{X} \in \mathcal{C}_{L}, \ \operatorname{rank}\left(\mathbf{X}\right)=1
	\end{aligned}
\end{equation}
where $\mathbf{X}=\mathbf{M}^{\mathrm{T}}\mathbf{M}$ is the covariance matrix of $\mathbf{M}$ and $\mathcal{C}_L$ defines the closed convex cone of completely positive (CP) matrices, whose definition is shown as below \cite{30}
%The set of CP matrices is defined as below \cite{30}
\begin{equation}
	%\begin{aligned}
		\mathcal{C}_{L}:=\left\{\mathbf{X} \in \mathcal{S}_{L}: \mathbf{X}=\mathbf{N} \mathbf{N}^{\mathrm{T}} \text { for some } \mathbf{N} \geqslant 0\right\},% \\
		%\mathcal{D}_{L}:=&\left\{\mathbf{X} \in \mathcal{S}_{L}: \mathbf{X} \succeq 0, \mathbf{X} \geqslant 0\right\}.
	%\end{aligned}
\end{equation}
where $\mathbf{X} \succeq 0$ denotes a positive semidifinite matrix and $\mathbf{X} \geqslant 0$ denotes an entry-wise nonnegative matrix.
\section{Proposed Pattern Design Algorithm}
%In this section, the matrix representation of the pattern channel is proposed, based on which the capacity maximized pattern design scheme is proposed.

In this section, we firstly show that the capacity maximization pattern design can be decomposed into the correlation modification process and the power allocation process. Subsequently, the sequential optimization framework and the closed-form power allocation scheme are proposed for each process.

%\subsection{Capacity Maximized Pattern Design}

Firstly, (11) can be further transformed into
\setcounter{equation}{14}
\begin{equation}
    \begin{aligned}
		\widetilde{\mathbf{H}} &=\mathbf{A}_{\mathrm{R}} \mathbf{\Lambda}\left(\mathbf{A}_{\mathrm{T}} \odot \mathbf{M}\right)^{\mathrm{H}} \\
		&=\sum_{i=1}^{L} \alpha_{i} \boldsymbol{a}_{\mathrm{R}, i}\left(\boldsymbol{a}_{\mathrm{T}, i} \odot \mathbf{M}(:, i)\right)^{\mathrm{H}} \\
		&=\sum_{i=1}^{L} \alpha_{i}p_i \boldsymbol{a}_{\mathrm{R}, i}\left(\boldsymbol{a}_{\mathrm{T}, i} \odot \widehat{\boldsymbol{m}}_{i}\right)^{\mathrm{H}}=\sum_{i=1}^{L} \alpha_{i}p_i\widehat{\mathbf{H}}_{i},
	\end{aligned}
\end{equation}
where $\widehat{\mathbf{H}}_{i}=\boldsymbol{a}_{\mathrm{R}, i}\left(\boldsymbol{a}_{\mathrm{T}, i} \odot \widehat{\boldsymbol{m}}_{i}\right)^{\mathrm{H}}$ is the $i$-th normalized modified subchannel. %and $p_i$ is the allocated power we distribute to the $i$-th modified subchannel.  %$\widehat{\boldsymbol{m}}_{i}=\frac{1}{p_i}\mathbf{M}(:,i)$ with $\left\|\widehat{\boldsymbol{m}}_{i}\right\|_2^2=N_t$ denotes the correlation modification vector of the $i$-th scattering path.
According to (15), the modification effect on the $i$-th scattering subchannel is accomplished via the correlation modification vector $\widehat{\boldsymbol{m}}_{i}=\frac{1}{p_i}\mathbf{M}(:,i)$ with $\left\|\widehat{\boldsymbol{m}}_{i}\right\|_2^2=N_t$, and the power allocation factor $p_i$.

From (15), the capacity maximization pattern design can be decomposed into two steps. Firstly, the correlation modification matrix $\widehat{\mathbf{M}}=\left[\widehat{\boldsymbol{m}}_1,\widehat{\boldsymbol{m}}_2, \ldots, \widehat{\boldsymbol{m}}_L\right]$ is designed to decrease the correlation level of the pattern channel. Subsequently, based on the modified channel structure, we propose a closed-form power allocation scheme to further distribute the communication resource wisely. In the following, the correlation modification problem is solved via the proposed sequential optimization framework firstly. Then we will further give an efficient power allocation scheme in a closed form.

\subsection{Correlation Modification Process}

%In order to quantify the correlation structure of the pattern channel, 
The covariance matrix $\widehat{\mathbf{G}} \in \mathbb{C}^{L \times L}$ of subchannels is given as (16) at the bottom of this page to describe the channel  correlation structure quantitatively.
%and it is obvious that $\widehat{\mathbf{G}}$ is a Hermitian matrix. 
Based on that, we define a correlation level indication vector in the following form:
\setcounter{equation}{16}
\begin{equation}
	\hat{\boldsymbol{g}}=\left[\sum_{j=2}^{L} \left|\widehat{\mathbf{G}}_{1,j}\right|^2, \sum_{j=1, j \neq 2}^{L} \left|\widehat{\mathbf{G}}_{2, j}\right|^2, \ldots, \sum_{j=1}^{L-1} \left|\widehat{\mathbf{G}}_{L, j}\right|^2\right]^{\mathrm{T}},
\end{equation}
%which quantifies the sum correlation effect of each subchannel, obtained from all the other subchannels.
whose $l$-th element describes the sum of the correlation between the $l$-th subchannel and all the other subchannels.
\subsubsection{Problem Formulation for SOF}
%To begin with, the correlation modification matrix $\widehat{\mathbf{M}}$ is initialized as $\mathbf{1}_{M \times N}$, which means that there is no modification effect for all subchannels. 
At the beginning, we initialize the correlation modification matrix $\widehat{\mathbf{M}}$ as $\mathbf{1}_{M \times N}$ so that all subchannels will not be modified in the first iteration.
In the $i$-th iteration, %we calculate the covariance matrix of each subchannel and further obtain the correlation indication vector $\hat{\boldsymbol{g}}$ in (17). 
after obtaining the correlation indication vector $\hat{\boldsymbol{g}}$, we further determine the the largest correlation level subchannel to be modified and denote the index as $n_i$. With $k \leq i-1$, the correlation between the $n_k$-th and the $n_i$-th normalized modified subchannel can be rewritten as

%After sorting the entries of $\hat{\boldsymbol{g}}$ in a descending order,
%According to $\hat{\boldsymbol{g}}$, we can obtain the subchannel with the largest correlation level that is to be updated within the current iteration, whose index is denoted by $n_i$. Without loss of generality, considering the correlation between the $n_k$-th and the $n_i$-th normalized modified channel with $k \leq i-1$, (16) can be simplified as
%\setcounter{equation}{37}
\begin{equation}
	\widehat{\mathbf{G}}_{n_{i}, n_{k}}=\rho_{n_{i}, n_{k}}^{\mathrm{R}} \boldsymbol{b}_{n_{i},n_{k}}^{\mathrm{T}} \widehat{\boldsymbol{m}}_{n_{i}},
\end{equation}
where %$\boldsymbol{b}_{n_{i}, n_{k}}$ is defined as
\begin{equation}
	\boldsymbol{b}_{n_{i}, n_{k}}(n)=\frac{1}{N_t}\widehat{\boldsymbol{m}}_{n_{k}}(n) e^{j 2 \pi \frac{d_{\mathrm{T}}}{\lambda}(n-1)\left(\sin \varphi_{n_{i}}-\sin \varphi_{n_{k}}\right)}
\end{equation}
and $\rho_{n_{i}, n_{k}}^{\mathrm{R}}=\frac{1}{N_r}\sum_{n=1}^{N_{r}} e^{j 2 \pi \frac{d_{\mathrm{R}}}{\lambda}(n-1)\left(\sin \theta_{n_{k}}-\sin \theta_{n_{i}}\right)}$ %represents the correlation coefficient between the $n_k$-th and the $n_i$-th receive steering vectors. 
quantifies the correlation effect at the receiver side.
Based on that, $\mathcal{P}_3$ is formulated to minimize the correlation between the $n_k$-th modified subchannel and the $n_i$-th updated subchannel, as shown in (20):
\begin{equation}
	\begin{aligned}
		\mathcal{P}_{3}: \min _{\widehat{\boldsymbol{m}}_{n_{i}}} \ & \left|\widehat{\mathbf{G}}_{n_{i}, n_{k}}\right|^{2} \\
		\text { s.t. }& \widehat{\boldsymbol{m}}_{n_{i}}^{\mathrm{T}} \widehat{\boldsymbol{m}}_{n_{i}}=N_{t}, \
		\widehat{\boldsymbol{m}}_{n_{i}} \geq 0,
	\end{aligned}
\end{equation}
where
\begin{equation}
	\begin{aligned}
		\left|\widehat{\mathbf{G}}_{n_{i}, n_{k}}\right|^{2}&=\left|\rho_{n_{i}, n_{k}}^{\mathrm{R}}\right|^{2}\left(\boldsymbol{b}_{n_{i}, n_{k}}^{\mathrm{T}} \widehat{\boldsymbol{m}}_{n_{i}}\right)^{\mathrm{H}} \boldsymbol{b}_{n_{i}, n_{k}}^{\mathrm{T}} \widehat{\boldsymbol{m}}_{n_{i}} \\
		&=\left|\rho_{n_{i}, n_{k}}^{\mathrm{R}}\right|^{2} \widehat{\boldsymbol{m}}_{n_{i}}^{\mathrm{T}}\left(\boldsymbol{b}_{n_{i}, n_{k}}^{*} \boldsymbol{b}_{n_{i}, n_{k}}^{\mathrm{T}}\right) \widehat{\boldsymbol{m}}_{n_{i}}\\
		&=\widehat{\boldsymbol{m}}_{n_{i}}^{\mathrm{T}}\operatorname{real}\left\{\left|\rho_{n_{i}, n_{k}}^{\mathrm{R}}\right|^{2}\boldsymbol{b}_{n_{i}, n_{k}}^{*} \boldsymbol{b}_{n_{i}, n_{k}}^{\mathrm{T}}\right\}\widehat{\boldsymbol{m}}_{n_{i}}\\
		&=\widehat{\boldsymbol{m}}_{n_{i}}^{\mathrm{T}}\mathbf{B}_{n_{k}}\widehat{\boldsymbol{m}}_{n_{i}},
	\end{aligned}
\end{equation}
and the first constraint is the power constraint that ensures $\left\|\widehat{\mathbf{H}}_{n_{i}}\right\|_{\mathrm{F}}^{2}=1$. 
%Considering that the real quadratic form will not be affected by the imaginary component of the conjugate symmetric coefficient matrix, the cost function of $\mathcal{P}_{3}$ can be simplified by defining
Considering that $\left|\widehat{\mathbf{G}}_{n_{i}, n_{k}}\right|$ defines a real quadratic with a conjugate symmetric coefficient matrix, we can simplify the optimization problem via 
$\mathbf{B}_{n_{k}}=\operatorname{real}\left\{\left|\rho_{n_{i}, n_{k}}^{\mathrm{R}}\right|^{2}\boldsymbol{b}_{n_{i}, n_{k}}^{*} \boldsymbol{b}_{n_{i}, n_{k}}^{\mathrm{T}}\right\}$.

In order to minimize the correlation between the $n_i$ subchannel and all the $(i-1)$ previously updated subchannels, the optimization problem in the $i$-th iteration is formulated as follows:
\begin{equation}
	\begin{aligned}
		\mathcal{P}_{4}: \min _{\widehat{\boldsymbol{m}}_{n_{i}}} \ &  \widehat{\boldsymbol{m}}_{n_{i}}^{\mathrm{T}}\left(\sum_{k=1}^{i-1} \mathbf{B}_{n_{k}}\right) \widehat{\boldsymbol{m}}_{n_{i}} \\
		\text { s.t. }& \widehat{\boldsymbol{m}}_{n_{i}}^{\mathrm{T}} \widehat{\boldsymbol{m}}_{n_{i}}=N_{t}, \
		\widehat{\boldsymbol{m}}_{n_{i}} \geq 0.
	\end{aligned}
\end{equation}
\subsubsection{Eigenvalue Decomposition Solution}
Considering the non-convexity introduced by the quadratic equality constraint, it is difficult to obtain the optimal solution of $\mathcal{P}_{4}$. In order to obtain a feasible solution, we propose an eigenvalue decomposition based algorithm. 
More specifically, based on $\mathbf{B}=\sum_{k=1}^{i-1} \mathbf{B}_{n_{k}}=\mathbf{U}\mathbf{\Sigma} \mathbf{U}^{\mathrm{T}}$, $\mathcal{P}_{4}$ can be further simplified as
\begin{equation}
	\begin{aligned}
		\mathcal{P}_{5}: \min _{\boldsymbol{m}} \ & \left(\mathbf{U}^{\mathrm{T}}\boldsymbol{m}\right)^{\mathrm{T}} \mathbf{\Sigma} \left(\mathbf{U}^{\mathrm{T}}\boldsymbol{m}\right) \\
		\text { s.t. } & \boldsymbol{m}^{\mathrm{T}} \boldsymbol{m}=N_{t}, \
		\boldsymbol{m} \geq 0,
	\end{aligned}
\end{equation}
where $\mathbf{U}$ is an $N_t \times N_t$ real orthogonal matrix and $\mathbf{\Sigma}$ contains the eigenvalues of $\mathbf{B}$ on the diagonal. Based on $\boldsymbol{w}=\mathbf{U}^{\mathrm{T}}\boldsymbol{m}$, $\mathcal{P}_{5}$ can be transformed into $\mathcal{P}_{6}$:
\begin{equation}
	\begin{aligned}
		\mathcal{P}_{6}: \min _{\boldsymbol{w}} \ & \boldsymbol{w}^{\mathrm{T}} \mathbf{\Sigma} \boldsymbol{w} \\
		\text { s.t. } & \boldsymbol{w}^{\mathrm{T}} \boldsymbol{w}=N_{t}, \
		\mathbf{U}\boldsymbol{w} \geq 0.
	\end{aligned}
\end{equation}

%Based on the fact that $\mathbf{\Sigma}$ is a diagonal matrix, a feasible solution to $\mathcal{P}_{14}$ can readily obtained as 
%Let $\boldsymbol{w}^{\star}=\sqrt{N_t} \mathbf{e}_{\mathrm{min}}$ where $\mathbf{e}_{\mathrm{min}}$ denotes a unit vector
%whose non-zero entry 
A sub-optimal solution of $\mathcal{P}_6$ can be obtained via $\boldsymbol{w}^{\star}=\sqrt{N_t} \mathbf{e}_{\mathrm{min}}$ where $\mathbf{e}_{\mathrm{min}}$ locates the smallest eigenvalue of $\mathbf{B}$
%corresponding to the location of the smallest eigenvalue of $\mathbf{B}$ 
%in $\operatorname{diag}\left\{\mathbf{\Lambda}\right\}$ 
\cite{34}. Based on that, a feasible solution to $\mathcal{P}_{4}$ can be obtained by
\begin{equation}
	\widehat{\boldsymbol{m}}_{n_{i}}^{\star}=\kappa\max \left\{\mathbf{U}\boldsymbol{w}^{\star}, \mathbf{0}\right\},
\end{equation}
where $\kappa=\sqrt{\frac{N_t}{\left\|\max \left\{\mathbf{U}\boldsymbol{w}^{\star}, \mathbf{0}\right\}\right\|_2^2}}$ is the power scaling factor for the satisfaction of $\left\|\widehat{\boldsymbol{m}}_{n_{i}}^{\star}\right\|_2^2=N_t$.

%Considering the non-convexity introduced by the quadratic equality constraint, it is difficult to obtain the optimal solution of $\mathcal{P}_{4}$. Fortunately, the mechanism of the SOF returns a point with a promising performance of the correlation modification process.

\subsection{Power Allocation Process}

Based on the optimized structure after the correlation modification process, an efficient power allocation scheme is needed to improve the channel quality. In this subsection, we firstly analyze the capacity maximized power allocation problem from the perspective of the singular value optimization to reveal the design principle. Then an efficient power allocation scheme with closed-form is proposed.

(8) can be transformed into
\begin{equation}
	\begin{aligned}
		%&-\log _{2} \operatorname{det}\left(\mathbf{I}_{N_{r}}+\frac{\rho}{N_{r}} \sum_{i=1}^{L}\sum_{j=1}^{L}\widetilde{\alpha}_i\widetilde{\alpha}_j^{\mathrm{H}}\mathbf{H}_i\mathbf{H}_{j}^{\mathrm{H}}\right) \\
		& -\log _{2} \operatorname{det}\left(\mathbf{I}_{N_{r}}+\frac{\rho}{N_{r}}\widetilde{\mathbf{H}}\widetilde{\mathbf{H}}^{\mathrm{H}}\right) \\
		=& -\log _{2} \prod_{i=1}^{r}\left(1+\frac{\rho}{N_{r}}\left|\sigma_i\left(\widetilde{\mathbf{H}}\right)\right|^2\right),
	\end{aligned}
\end{equation}
where $\sigma_i\left(\widetilde{\mathbf{H}}\right)$ denotes the $i$-th singular value of $\widetilde{\mathbf{H}}$ while $r$ is the number of non-zero singular values. Based on that, the power allocation problem can be formulated as:
%where $r=\operatorname{rank}\left(\widetilde{\mathbf{H}}\right)$ is the rank of $\widetilde{\mathbf{H}}$ and $\sigma_i\left(\widetilde{\mathbf{H}}\right)$ denotes the $i$-th singular value of $\widetilde{\mathbf{H}}$. Based on (15) and (26), the power allocation problem can be generalized into:
\begin{equation}
	\begin{aligned}
		\mathcal{P}_{7}: \min _{\left\{x_i\right\}} \ &-\prod_{i=1}^{r}\left(1+\gamma x_i\right) \\
		\text { s.t. } & \sum_{i=1}^{r}x_i-N_{r} N_{t} \leq 0, \ 
		%&x_i=\left|\lambda_i\left(\sum_{l=1}^{L}\tilde{\alpha}_{l} \mathbf{H}_{l}\right)\right|^2
		 \gamma=\frac{\rho}{N_r},
	\end{aligned}
\end{equation}
where
\begin{equation}
	x_i=\left|\sigma_i\left(\widetilde{\mathbf{H}}\right)\right|^2=\left|\sigma_i\left(\sum_{l=1}^{L}\tilde{\alpha}_{l} \widehat{\mathbf{H}}_{l}\right)\right|^2
\end{equation}
and $\tilde{\alpha}_l=\alpha_lp_l$ is the channel gain of the $l$-th path after the power allocation process.

%Without considering (30), the optimal solution $\left\{x_i^{\star}\right\}$ of $\mathcal{P}_7$ can be obtained easily through the Karush-Kuhn-Tucker (KKT) conditions. However, the non-convexity introduced by (30) generally excludes $\left\{x_i^{\star}\right\}$ from the feasible region.
%Although the non-convexity introduced by (28) excludes the optimal solution $\left\{x_i^{\star}\right\}$ of $\mathcal{P}_7$ from the feasible region,
Although the non-convexity introduced by (28) prevent the optimal solution of $\mathcal{P}_7$, (27) reveals the relationship between the pattern channel and its eigen-subchannels.
%(27) and (28) exhibit a clear physical interpretation that the capacity of the channel is equivalent to the sum capacity of the independent eigen-subchannels. %Each eigenvalue is the quantitative description of how much communication resource can be distributed to the corresponding eigen-subchannel. 
%Similar to the principle of water-filling, a "good" channel means that the singular values of the channel matrix are distributed as even as possible, and vice verse. Based on that, the power we distribute to each subchannel should be proportional to its independence. The closed-form power allocation (CFPA) scheme is given by
According to the water-filling principle, the power we distribute to each subchannel should be proportional to its independence so that the singular values of the channel matrix can be distributed as even as possible. The closed-form power allocation (CFPA) scheme is given by
\begin{equation}
    %\boldsymbol{w}=\frac{\max\left(\hat{\boldsymbol{g}}\right)/\hat{\boldsymbol{g}}}{\left\|\max\left(\hat{\boldsymbol{g}}\right)/\hat{\boldsymbol{g}}\right\|^2},
    %\boldsymbol{w}=\frac{\max\left(\hat{\boldsymbol{g}}\right)/\hat{\boldsymbol{g}}}{\mathbf{1}^{\mathrm{T}}},
    \hat{w_l}=\frac{\max\left(\hat{\boldsymbol{g}}\right)}{\hat{g}_l},
\end{equation}
where $\boldsymbol{w}=\frac{\hat{\boldsymbol{w}}}{\mathbf{1}^{\mathrm{T}}\hat{\boldsymbol{w}}}$ is the power proportion vector. The power scaling factor $\delta$ for the satisfaction of the channel power constraint can be defined as follows:
%\begin{equation}
%	\begin{aligned}
%		&\left\|\delta\sum_{l=1}^{L}w_l\widehat{\mathbf{H}}_l\right\|_{\mathrm{F}}^2=N_tN_r \\
%		\Leftrightarrow&{\delta}^2\mathrm{Tr}\left(\left(\sum_{l=1}^{L}w_l\widehat{\mathbf{H}}_l\right)^{\mathrm{H}}\left(\sum_{l=1}^{L}w_l\widehat{\mathbf{H}}_l\right)\right)=N_tN_r ,\\
%		\Leftrightarrow&\delta=\sqrt{\frac{N_tN_r}{\mathrm{Tr}\left(\left(\sum_{l=1}^{L}w_l\widehat{\mathbf{H}}_l\right)^{\mathrm{H}}\left(\sum_{l=1}^{L}w_l\widehat{\mathbf{H}}_l\right)\right)}}.
%	\end{aligned}
%\end{equation}

\begin{equation}
%	\begin{aligned}
%		&\left\|\delta\sum_{l=1}^{L}w_l\widehat{\mathbf{H}}_l\right\|_{\mathrm{F}}^2=N_tN_r \\
%		\Leftrightarrow&{\delta}^2\mathrm{Tr}\left(\left(\sum_{l=1}^{L}w_l\widehat{\mathbf{H}}_l\right)^{\mathrm{H}}\left(\sum_{l=1}^{L}w_l\widehat{\mathbf{H}}_l\right)\right)=N_tN_r ,\\
		\delta=\sqrt{\frac{N_tN_r}{\mathrm{Tr}\left(\left(\sum_{l=1}^{L}w_l\widehat{\mathbf{H}}_l\right)^{\mathrm{H}}\left(\sum_{l=1}^{L}w_l\widehat{\mathbf{H}}_l\right)\right)}}.
%	\end{aligned}
\end{equation}

Let $\left|\tilde{\alpha}_l\right|=w_l\delta$ for all $l=1,2,\ldots,L$. Then the allocated power factor of the $l$-th path is given by
\begin{equation}
	p_l=\frac{\left|\tilde{\alpha}_l\right|}{\left|\alpha_l\right|}=\frac{w_l\delta}{\left|\alpha_l\right|}.
\end{equation}

The overall algorithm for the pattern design algorithm is summarized in Algorithm 1.
\begin{algorithm}[htbp]
	\caption{Sequential Optimization Framework (SOF)}
	\label{alg2}
	\begin{algorithmic}[1]
		\REQUIRE $\mathbf{A}_{\mathrm{R}}$, $\mathbf{A}_{\mathrm{T}}$, $\mathbf{\Lambda}$
		\ENSURE $\mathbf{M}$
		\STATE Initialize $\mathcal{I}=\emptyset$ and $\widehat{\mathbf{M}}=\mathbf{1}_{N_t \times L}$;
		\STATE Calculate $\mathbf{H}_i=\boldsymbol{a}_{\mathrm{R},i}\boldsymbol{a}_{\mathrm{T},i}^\mathrm{H}$; Obtain $\mathcal{S}=\left\{\mathbf{H}_{i} \mid i=1,2, \ldots, L\right\}$;
		%\STATE Calculate $\widehat{\mathbf{G}}^{(1)}$ based on (16); Calculate $\widehat{\boldsymbol{g}}^{(1)}$ based on (17);
		\STATE Calculate $\widehat{\mathbf{G}}^{(1)}$ and $\widehat{\boldsymbol{g}}^{(1)}$ based on (16) and (17);
		%\STATE Sort $\boldsymbol{g}^{(1)}$ in ascending order and obtain the index vector $\boldsymbol{r}^{(1)}$;
		\STATE Find $n_1$ such that $\widehat{\boldsymbol{g}}_{n_1}^{(1)}=\operatorname{max}\left\{\widehat{\boldsymbol{g}}^{(1)}\right\}$; Stack $\mathcal{I}=[\mathcal{I},n_1]$;
		%\STATE Obtain $n_1=\boldsymbol{r}^{(1)}(1)$; Stack $\mathcal{I}=[\mathcal{I},n_1]$;
		\FOR{$i=2:L$}
		%\STATE Remove $\mathcal{I}$ from $\boldsymbol{r}^{(i)}$; Obtain $n_i=\boldsymbol{r}^{(i)}(1)$ and  $\boldsymbol{s}^{(i)}=[\mathcal{I},\boldsymbol{r}^{(i)}]$;
		\STATE Remove $\widehat{\boldsymbol{g}}_j^{\left(i-1\right)}$ from $\widehat{\boldsymbol{g}}^{\left(i-1\right)}$ and obtain $\widehat{\boldsymbol{g}}'$, ${\forall}j \in \mathcal{I}$;
		\STATE Find $n_i$ such that $\widehat{\boldsymbol{g}}_{n_i}^{(i)}=\operatorname{max}\left\{\widehat{\boldsymbol{g}}'\right\}$; Stack $\mathcal{I}=\left[\mathcal{I},n_i\right]$;
		\STATE Initialize $\mathbf{B}_i=\boldsymbol{0}_{L \times L}$;
		\FOR{$k=1:(i-1)$}
		\STATE Obtain $n_k=\mathcal{I}(k)$;
		\STATE Calculate $\boldsymbol{b}_{n_i,n_k}$ based on (19); 
		\STATE Obtain $\mathbf{B}_{n_k}=\boldsymbol{b}_{n_i,n_k}^{*}\boldsymbol{b}_{n_i,n_k}^{\mathrm{T}}$;
		\STATE Update $\mathbf{B}_i=\mathbf{B}_i+\mathbf{B}_{n_k}$;
		\ENDFOR
		\STATE Solve $\mathcal{P}_{4}$ and obtain $\widehat{\boldsymbol{m}}_{n_i}^{\star}$; Update $\widehat{\mathbf{M}}(:,n_i)=\widehat{\boldsymbol{m}}_{n_i}^{\star}$;
		%\STATE Update $\widehat{\mathbf{M}}(:,n_i)=\kappa\max{\left\{\widehat{\boldsymbol{m}}_{n_i},\boldsymbol{0}_{L \times 1}\right\}}$;
		%\STATE Update $\widehat{\mathbf{M}}(:,n_i)=\widehat{\boldsymbol{m}}_{n_i}^{\star}$;
		\STATE Update $\widehat{\mathbf{H}}_{n_i}=\boldsymbol{a}_{\mathrm{R},n_i}\left(\boldsymbol{a}_{\mathrm{T},n_i} \odot \widehat{\mathbf{M}}(:,n_i)\right)^{\mathrm{H}}$;
		%\STATE Calculate $\widehat{\mathbf{G}}^{(i)}$ based on (16); Calculate $\widehat{\boldsymbol{g}}^{(i)}$ based on (17);
		\STATE Calculate $\widehat{\mathbf{G}}^{(i)}$ and $\widehat{\boldsymbol{g}}^{(i)}$ based on (16) and (17);
		%\STATE Sort $\boldsymbol{g}^{(i)}$ in ascending order and update the index vector $\boldsymbol{r}^{(i)}$;
		\ENDFOR
		\STATE Obtain $\widehat{\mathcal{S}}=\left\{\widehat{\mathbf{H}}_i \mid i=1,2, \ldots, L\right\}$;
		%\STATE Calculate $\widehat{\mathbf{G}}$ based on (16); Calculate $\widehat{\boldsymbol{g}}$ based on (17);
		\STATE Calculate $\widehat{\mathbf{G}}$ and $\widehat{\boldsymbol{g}}$ based on (16) and (17);
		%\STATE Solve $\mathcal{P}_5$ and obtain $\boldsymbol{p}$;
		\STATE Obtain $p_i$ based on (31);
		\STATE Output $\mathbf{M}=\widehat{\mathbf{M}}\mathrm{diag}\left\{p_i\right\}$;
		
	\end{algorithmic}
\end{algorithm}

\section{Numerical Results}
Numerical results are presented in this section. With $N_{\mathrm{cl}}$ scattering clusters and $N_{\mathrm{ray}}$ scattering pathes in each cluster, (1) can be expanded into
\begin{equation}
	\mathbf{H}=\sum_{i=1}^{N_{\mathrm{cl}}}\sum_{j=1}^{N_{\mathrm{ray}}}\alpha_{i,j}\boldsymbol{a}_{\mathrm{R}}\left(\theta_{i,j}\right)\boldsymbol{a}_{\mathrm{T}}^{\mathrm{H}}\left(\varphi_{i,j}\right).
\end{equation}

\begin{figure}[t]
	\centering
	\includegraphics[width=3in]{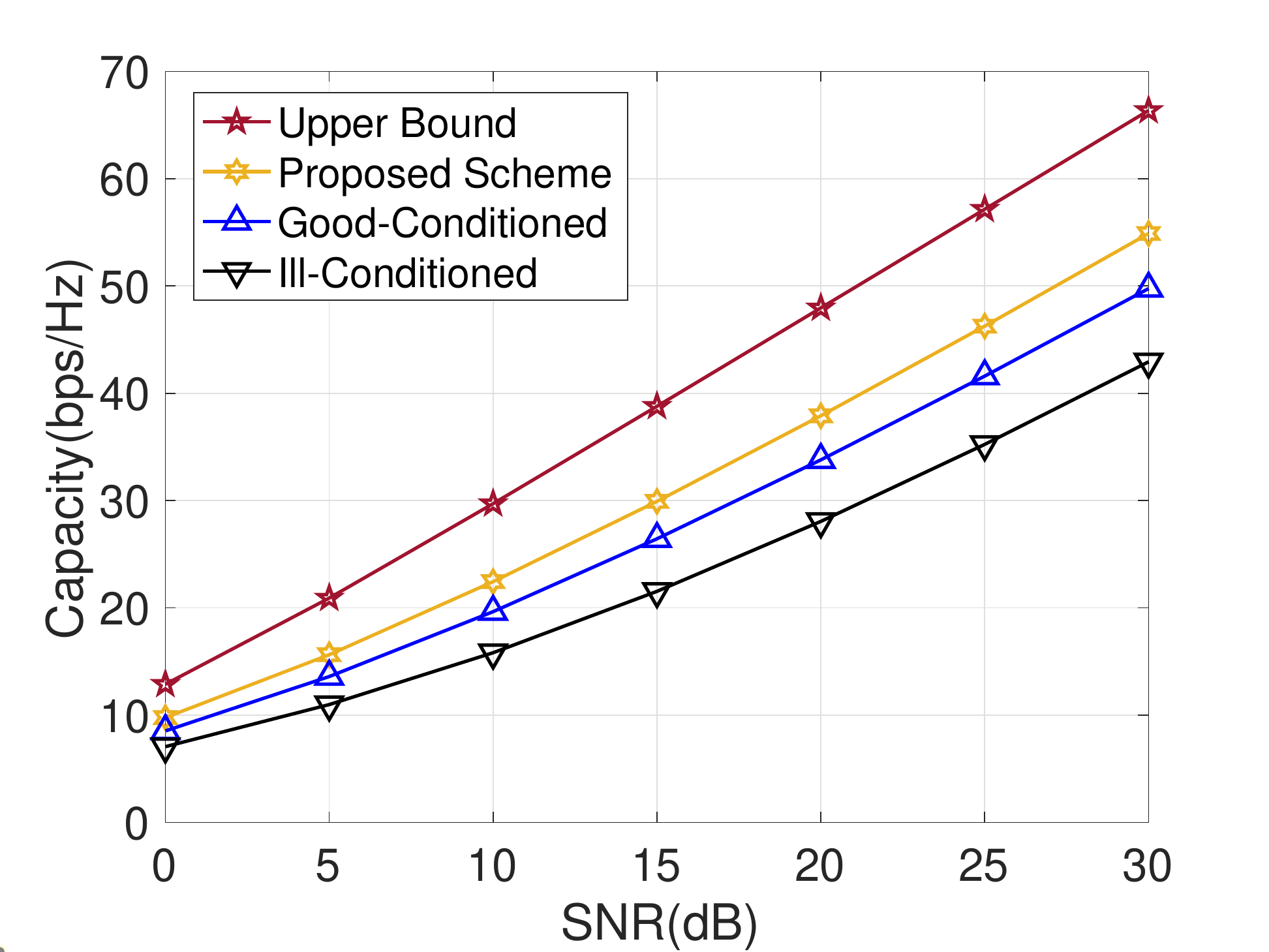}
	\caption{Capacity v.s. transmit SNR, $N_t=32$, $N_r=8$, $N_{\mathrm{cl}}=10$, $N_{\mathrm{ray}}=8$, physical and pattern channels. }
	\label{structure}
\end{figure}
\begin{figure}[t]
	\centering
	\includegraphics[width=3in]{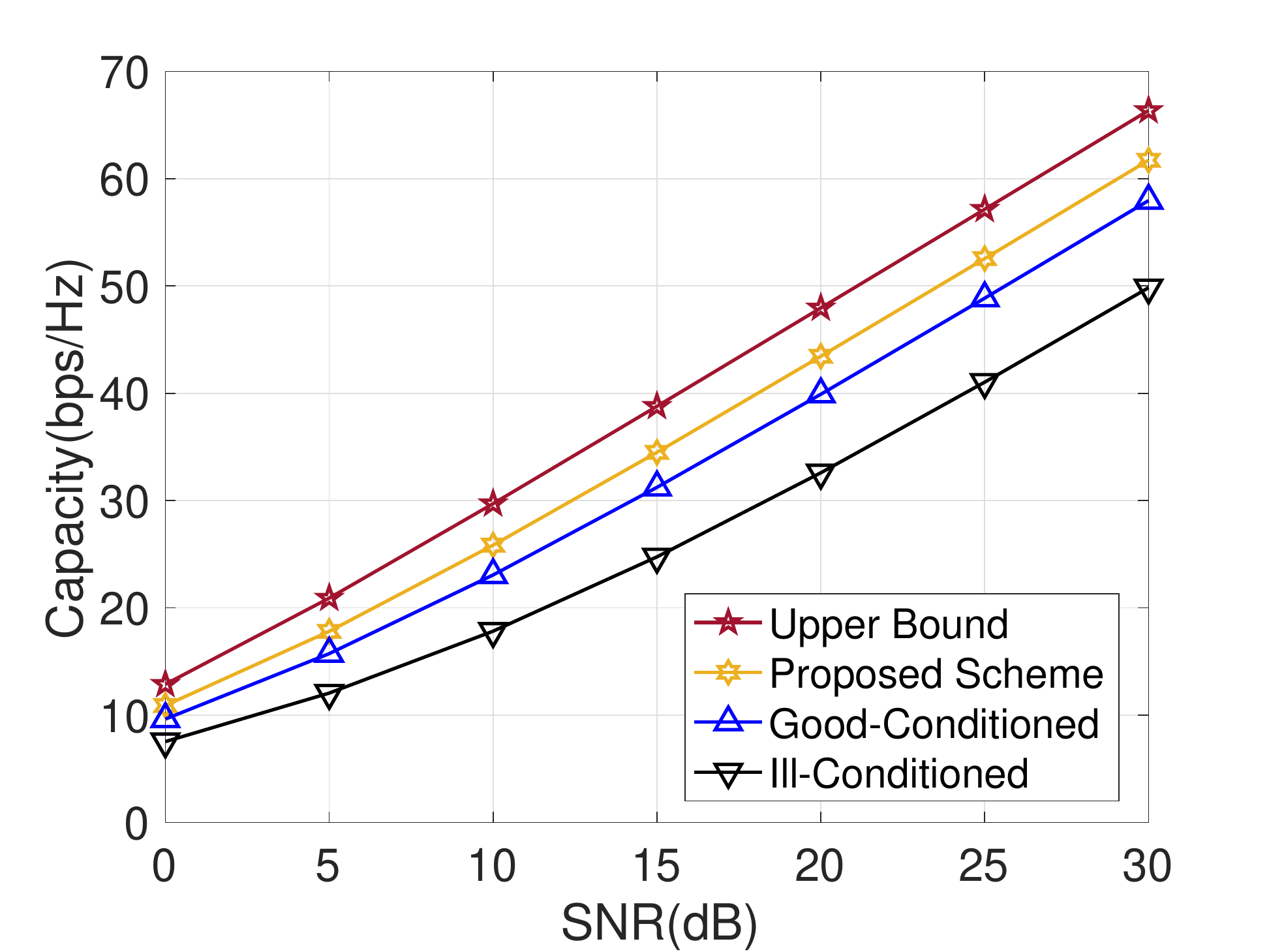}
	\caption{Capacity v.s. transmit SNR, $N_t=32$, $N_r=8$, $N_{\mathrm{cl}}=20$, $N_{\mathrm{ray}}=8$, physical and pattern channels. }
	\label{structure}
\end{figure}

We consider that $\alpha_{i,j}$ follows $\mathcal{C} \mathcal{N}\left(0, \sigma_{i}^{2}\right)$, where $\sigma_{i}^{2}$ represents the average power of the $i$-th cluster with $\sum_{i=1}^{N_{\mathrm{cl}}} \sigma_{i}^{2}=\gamma$.
We introduce $\gamma$ as the normalization factor so that the power constraint of the channel matrix $\mathbb{E}\left\{\left\|\mathbf{H}\right\|_{\mathrm{F}}^{2}\right\}=N_{r} N_{t}$ can be held.
%and $\gamma$ is the normalization parameter to ensure that $\mathbb{E}\left\{\left\|\mathbf{H}\right\|_{\mathrm{F}}^{2}\right\}=N_{r} N_{t}$. 
$\theta_{i,l}$ is uniformly distributed with mean $\theta_{i}$ and a standard deviation $\xi$, and $\varphi_{i,l}$ is uniformly distributed with mean $\varphi_{i}$ and the same standard deviation $\xi$. 
%Considering that the distribution of $\left\{\sigma_{i}^{2}\right\}$ determines the condition number of the channel matrix, 
%For the good-conditioned channel, $\left\{\sigma_{i}^{2}\right\}$ follows the normal distribution so that the channel condition number will be small. While for the ill-conditioned channel, we set $\sigma_{1}^{2}:\sigma_{2}^{2}:\sigma_{3}^{2}:\sigma_{4}^{2}: \ldots : \sigma_{N_{\mathrm{cl}}}^{2}=100:50:50:1: \ldots :1$ to obtain a large condition number.
In order to distinguish the physical channel quality, we define the good-conditioned channel whose $\left\{\sigma_{i}^{2}\right\}$ are the normal random variables to obtain a relatively small channel condition number, and the ill-conditioned channel with $\sigma_{1}^{2}:\sigma_{2}^{2}:\sigma_{3}^{2}:\sigma_{4}^{2}: \ldots : \sigma_{N_{\mathrm{cl}}}^{2}=100:50:50:1: \ldots :1$ so that the channel condition number will be large.
All the results are obtained by averaging over 1000 Monte Carlo simulations. Unless otherwise stated, $N_t=32$, $N_r=8$, $\xi=3^{\circ}$, and the half-wavelength antenna spacing is held for all simulations. Both $\theta_{i}$ and $\varphi_{i}$ are  uniformly distributed in the range of $\left[-\pi/2, \pi/2\right]$. We adopt the ideal channel $\mathbf{H}_{\mathrm{opt}}=\sqrt{N_t}\mathbf{I}_{N_r \times N_t}$ as the theoretical performance upper bound.

Fig. 2 compares the performance of the proposed pattern design scheme with the theoretical ideal channel, the good-conditioned channel and the ill-conditioned channel, when $N_{\mathrm{cl}}=10$. We can see that the proposed pattern design scheme can improve the capacity performance compared with the physical channel, which validates that the quality of the channel can be significantly improved by the pattern modification effect. Meanwhile, the gap between the proposed scheme and the theoretical upper bound reveals that the inability of phase adjustment prevents the further performance improvement of the PE-MIMO system.

Fig. 3 shows the analogous simulation results of Fig. 2 when $N_{\mathrm{cl}}=20$. Compared with Fig. 2, 
%the achievable capacity of the physical channel in good condition is better for the additional subchannels. 
the additional subchannels provide more spatial DoFs so that the achievable capacity of the physical channel in good condition is improved.
%Moreover, the performance gap between the proposed pattern design scheme and the theoretical upper bound becomes smaller, which shows that the modification ability is proportional to the number of clusters.
Moreover, the smaller gap between the performance of the proposed pattern design scheme and the theoretical upper bound reveals that the increasing scattering clusters provide additional DoFs which improves the pattern modification effect.

\section{Conclusion}

In this paper, we study a wireless communication system towards EIT that is able to configure the data transmission and electro-magnetic field distribution via the concept of PR-MIMO. Based on the matrix representation of PR-MIMO we propose, it is shown that the effect of radiation reconfigurability can be regarded as an additional gain on the corresponding propagation directions, and the capacity maximization pattern design problem is further formulated. We further decompose the optimal pattern design into the correlation modification process and the power allocation process, and propose an efficient design algorithm. More specifically, the channel correlation structure is optimized via the correlation modification matrix in the correlation modification process. Subsequently, based on the optimized subchannels, the transmission power is redistributed wisely in the power allocation process. Based on that, for the correlation modification process, we propose a sequential optimization framework with eigenvalue decomposition. For the power allocation process, a closed-form scheme is designed to improve the channel quality. Numerical results validate the superiority of PR-MIMO over traditional MIMO systems as well as the effectiveness of proposed algorithms. The future work will discuss the combination between PR-MIMO and constructive interference precoding \cite{39}.
\bibliographystyle{IEEEtran}
\bibliography{IEEEreference}
\end{document}